\pdfoutput=1
\documentclass[conference]{IEEEtran}

\ifCLASSINFOpdf
\else
\fi

\usepackage{url}

\usepackage{amsfonts}
\usepackage{amsmath}
\usepackage{amssymb}

\usepackage{graphicx}
\graphicspath{{images/}}
\DeclareGraphicsExtensions{.pdf}

\begin{document}
\title{Using Machine Learning to Detect Noisy Neighbors in 5G Networks}





%
\author{\IEEEauthorblockN{Udi Margolin\IEEEauthorrefmark{1},
Alberto Mozo\IEEEauthorrefmark{2},
Bruno Ordozgoiti\IEEEauthorrefmark{2}, 
Danny Raz\IEEEauthorrefmark{1},
Elisha Rosensweig\IEEEauthorrefmark{1} and
Itai Segall} 
\IEEEauthorblockA{\IEEEauthorrefmark{1}Bell Labs Israel}
\IEEEauthorblockA{\IEEEauthorrefmark{2}Universidad Polit\'{e}cnica de Madrid\\
Email: bruno.ordozgoiti@etsisi.upm.es, danny.raz@nokia.com}
}


\maketitle

\begin{abstract}
5G networks are expected to be more dynamic and chaotic in 
their structure than current networks. With the advent of Network
Function Virtualization (NFV), Network Functions (NF) will no longer
be tightly coupled with the hardware they are running on, which poses
new challenges in network management. Noisy neighbor is a term
commonly used to describe situations in NFV infrastructure where an
application experiences degradation in 
performance due to the fact that some of the resources it needs are
occupied by other applications in the same cloud node. These situations
cannot be easily identified using straightforward approaches, which
calls for the use of sophisticated methods for NFV infrastructure
management. In this paper 
we demonstrate how Machine Learning (ML) techniques can be used to
identify such events. Through experiments using data collected at real
NFV infrastructure, we show that standard models for automated
classification can detect the noisy neighbor phenomenon with an
accuracy of more than 90\% in a simple scenario. 

\end{abstract}


%
\IEEEpeerreviewmaketitle

\section{Introduction}
\label{sec:intro}


5G networks are expected to be inherently more dynamic and chaotic in
their structure than current networks. With the advent of Network
Function Virtualization (NFV), Network Functions (NF) will no longer
be tightly coupled with the hardware they are running on. Instead,
when considering network design and management a decade into the
future, it is believed that many NFs will be virtualized, and thus
able to be deployed, scaled and migrated with ease and speed unheard
of in today’s networks. As a result, while the topology of the
physical network infrastructure shall evolve on a slow time scale
similar to that of current networks, the dynamics of the
virtual/logical topology will undergo a dramatic change and be allowed
to evolve at much faster speeds, in service of customer needs and
demands. 

When considering network management in such an environment, achieving
the degree of reliability and stability that is to be expected of 5G
becomes a challenge. Specifically, when the network topology is in
constant flux, it is more difficult to make decisions regarding where
to locate services, ensure the SLA a provider is committed to, and
perform Root Cause Analysis (RCA) on system faults.  
One specific challenge in this domain is the problem of allocating
network resources in a dynamically changing environment like NFV
\cite{etsinfv}.   This challenge relates to the problem of service
demand prediction and provisioning which allows the network to resize and
resource itself, using virtualization, to serve predicted demand
according to parameters such as location, time and specific service
demand from specific users or user groups. 

The more specific issues we will address in this paper is identifying
the state of a service and the reason for this state. More
specifically we want to identify that the service is facing
difficulties and identify the cause of these difficulties. This is a
critical building block in addressing use cases similar to the
“Optimized Services in Dynamic Environments” use case described in
\cite{buda2016machine}. 
Noisy neighbor is a term commonly used to describe situations in cloud
computing where an application (or a VM) experiences degradation in
performance due to the fact that some of the resources it needs are
occupied by other applications (VMs) in the same cloud node.  Note
that in general there should be a clear separation between the
different tenants on the same physical machine, but in practice this
isolation of the virtual machines is far from perfect and many
resources such as internal networking resources and memory access
resources are shared at some level. In this paper we demonstrate how
Machine Learning (ML) techniques can be used to identify such events.
We show that using standard models for automated classification, noisy neighbors
can be detected in a simple scenario with more than 90\% accuracy.
As said this is a critical building block in creating flexible
reliable orchestration mechanisms for 5G networks as once identified,
the system can resolve the performance issue by migrating one of the
tenants to another machine or by allocating more resources.  The rest
of the paper is structured as folows: section \ref{sec:problem}
describes the problem setting in detail. Section \ref{sec:ml}
describes the employed machine learning models. Section \ref{sec:exps}
provides experimental results and finally section \ref{sec:conclusion}
outlines future work and offers concluding remarks.

\section{Problem setting}
\label{sec:problem}
This testbed environment consists of a total of five high performance
HP servers (each having 12 cores) with a proprietary management system
running on top of an OpenStack cloud \cite{openstack}. Two of the servers are
dedicated to the management processes, while the other three are
available as compute machines. 

An open source voice-over-IP (VoIP) application, Asterisk \cite{asterisk}, is
deployed on one of the compute machines, in a single virtual machine
utilizing one core and 1024 MB of memory. A corresponding traffic
generator, SipP \cite{sipp} is deployed in a second machine, and configured to
log in and log out of the server, and initiate calls to a line in
which the server is playing music-on-hold. The traffic generator is
configured in a manner that generates constant reasonable load on the
server, e.g. 40\% CPU utilization.  

Another application that was considered is a Fibonacci server – a
specially-tailored application, designed to mimic Virtual Network
Functions in behaviour, but be simple enough for experimentation and
analysis. The server computes elements in the Fibonacci series on
demand. A client may request the index of the element to be computed.
Different implementations of the server are considered, to examine the
noisy neighbour effect on them. 

A noisy neighbour is defined as (one or more) virtual machines (VMs)
sharing resources with the server under test, thus affecting its
performance. Since both of the servers under test are CPU intensive,
we focused on CPU noise. Noise was generated using the "stress" Linux
application. In its CPU-intensive setting, it continuously computes
square-roots of numbers in order to generate load on CPUs. Noise is
therefore simulated by further noise generation virtual machines, all
launched on the same server as the VoIP server. We experimented with
both a single large VM occupying all physical CPUs, as well as
multiple small VMs, each occupying only one or two cores. 
During experimentations, the following metrics were collected:
\begin{itemize}
\item 	CPU utilization of the server VM
\item 	In-bound network traffic of the server VM
\item 	Out-bound network traffic of the server VM
\item 	CPU utilization of the noise VM(s). The purpose of this metric
  is to act as an indicator of whether noise exists or not, for the ML
  training period
\end{itemize}
Monitoring is performed using Openstack Ceilometer, and the data is collected by the proprietary management system into a DB for further analysis and extraction.
Experimentations differ in the following attributes:
\begin{itemize}
\item 	Server under test: Asterisk vs. Synthetic
\item 	Amount of noise: One VM utilizing all 24 cores vs. several
  (between 18 and 24) utilizing a single core each. 
\item 	Amount of traffic generated (and respectively amount of CPU
  utilization in the server) 
\end{itemize}

\section{Machine learning methods}
\label{sec:ml}
The problem of determining whether or not the behavior of a virtual
machine is being caused by the presence of a noisy neighbor is
non-trivial. Based on the metrics described above, a simple
thresholding approach or a set of rules would not suffice. 

Therefore, to address this problem we propose the use of machine
learning methods. Machine learning has proved successful in a wide
variety of domains, such as computer vision
\cite{salakhutdinov2009deep} \cite{krizhevsky2012imagenet}, bioinformatics \cite{yang2010review}, natural
language processing
\cite{morin2005hierarchical}\cite{sundermeyer2012lstm} and others. A key
advantage of machine 
learning is that the same models can be applied to completely
different domains provided that the data are transformed into an
adequate representation.

In this paper we address the problem of determining whether or not a
noisy neighbor situation is taking place, which can be modelled in
machine learning as a classification problem. Hence, we propose to
employ two well-known classifiers: support vector machines and random
forests. 

\subsection{Support vector machines}
Given a set of data points represented as real-valued vectors
corresponding to two different classes, support
vector machines (SVM) \cite{cortes1995support} pose the problem of finding a maximum-margin
separating hyperplane.  It can be shown that this is equivalent to
minimizing the norm of the vector orthogonal to said hyperplane,
constraining the data points of different classes to be on different
sides and at a minimum distance from it. Formally, this can be formulated as follows.

Given a set of data points $\{x_1, \dots, x_n\}$ and corresponding
labels $\{y_1, \dots, y_n\}$, where $x_i \in \mathbb{R}^{d}$ and $y_i
\in \{-1,1\}$, $i=1, \dots, n$, for some $d \in  \mathbb{N}$, find

\begin{equation}
\label{eq:svm_primal}
\begin{aligned}
& \underset{w}{\text{min}} & &  \frac{1}{2}\|w\|^2 + C \sum_i{\xi_i}  \\
& \text{s.t.} & & y_i(wx_i) \geq 1 - \xi_i \mbox{ },  i = 1, \dots, n \\
& & & \xi_i \geq 0 \mbox{
    }  i = 1, \dots, n
\end{aligned}
\end{equation}

Here, we introduce slack variables $\xi_i$, which allow for violations
of the separation constraint. This formulation is often referred to as
\textit{soft-margin} SVM. The variable $C \in \mathbb{R}$ is a
hyperparameter that allows us to control the amount of penalization
incurred by these violations.

In the case of support vector machines, strong duality holds
\cite{boyd2004convex}. Therefore, a Lagrangian 
can be formulated such that its maximum value provides the solution to
our problem. Optimizing with respect to the corresponding variables
yields the following formulation.

\begin{equation}
\begin{aligned}
& \underset{\alpha}{\text{max}} & &  \sum_i\alpha_i -
  \frac{1}{2}\sum_i\sum_j\alpha_i\alpha_j y_i y_j\langle x_i,x_j\rangle  \\
& \text{s.t.} & & \sum_iy_i\alpha_i = 0 \\
& & & 0 \leq \alpha_i \leq C \mbox{
    }  i = 1, \dots, n
\end{aligned}
\end{equation}
Here, $\alpha_i$, $i=1, \dots, n$ are the Lagrange multipliers
corresponding to the inequality constraint of the primal formulation.

The sole dependence of this formulation on the inner products allows for
the use of a technique often referred to as the \textit{kernel 
  trick} \cite{boser1992training}. Kernel functions implicitly map vectors to a higher
dimensional space and compute their inner product. This enables
learning separating hyperplanes for non-separable data.

\subsection{Random forests}
Random forests \cite{breiman2001random} are a learning model that combines decision trees with
the concept of bagging \cite{breiman1996bagging} and random feature selection. Random
forests are widely popular because of their simplicity, efficiency and
effectiveness. 

Random forests function by combining bagging, or bootstrap
aggregation, with decision trees. Specifically a number of random samples
with replacement is drawn from the data set and a decision tree is
trained for each one of them. The final decision for classifying an unseen
data instance is taken by majority vote of the trained decision trees.
Decision trees, in turn, are built by constructing a hierarchy of
binary split nodes. The criterion to choose the splitting rule for
each node is the maximization of information gain based of some
measure of entropy.

In addition to the bagging process, random forests also employ random 
subsets of features. In our case, though, due to the low number of
features we limit the model to bagged decision trees.

\section{Experimental results}
\label{sec:exps}
In this section, we describe the experiments carried out in order to
assess the validity of machine learning as a method for detecting
noisy neighbors, as well as the results obtained. 
\subsection{Data} We collected the Ceilometer metrics as described in 
section \ref{sec:problem}, running around 100 experiments of the
system described above. The metrics were collected each 10 seconds approximately. 
These measurements were aggregated into 30 second periods in order to
avoid the impact of missing values or irregularities in the sampling
frequency. Each of the resulting data points thus represents the
average CPU load, inbound and outbound bandwidth of the monitored
machine. The corresponding binary label determines whether or not the
noisy neighbor was inflicting load during that period. The resulting
data set is comprised of 9169 data instances, out of which 3088
correspond to noisy neighbors. 

As stated above, detecting noisy neighbors from the available data is
non-trivial. To illustrate this fact, Table \ref{tab:corrmic} shows
two different measurements of the relationship between the collected
features and the noise value. The first row shows the correlation coefficient
between each of the different features and the CPU load of the machine
that creates noise. The second row shows the maximal information
coefficient (MIC) \cite{reshef2011detecting}. It is apparent that none
of the available features is sufficiently reliable as a predictor on
its own, suggesting that the noisy neighbor phenomenon manifests
itself in a complex manner.

\begin{table}[!t]
\centering
\begin{tabular}{|c|c|c|c|}
\hline
\multicolumn{4}{|c|}{Correlation and MIC} \\
\hline
& CPU & BW in & BW out \\
\hline
Correlation & 0.125 & 0.477 & -0.092
\\ \hline
MIC & 0.371 & 0.341 & 0.306
\\ \hline
\end{tabular}
\vspace{1em}
\caption{Correlation and maximal information coefficient between the
  collected features and the noise.} 
\label{tab:corrmic}
\end{table}

\subsection{Experiments}
We train the two models described in section \ref{sec:ml} to
automatically classify data instances as corresponding to a noisy
neighbor situation or not. The input data is standardized to zero mean
and unit variance. In both cases we follow a 10-fold 
cross validation process. We now describe the specific training
procedures for each of the models. 
\subsubsection{Support vector machines} We train $\ell 1$
soft-margin models with the sequential minimal optimization algorithm
\cite{platt1998sequential} using a Gaussian kernel. The Gaussian kernel can be defined as
$K(u, v) = e^{-\gamma \|u-v\|^2}$, where $\gamma$ is a
hyperparameter. We train models using different values of $C$. A
quadratic expansion of the features is employed, i.e. each data
instance $(x_1, x_2, x_3)$ is transformed into $(x_1, x_2, x_3, x_1^2,
x_2^2, x_3^2, x_1x_2, x_1x_3, x_2x_3)$.
\subsubsection{Random forests} As stated above, given the small number
of features we limit the model to bagged decision trees employing all
features. Gini's diversity index is employed for splitting nodes.

\subsubsection{Results}
In our experiments, both models attained acceptable levels of
classification performance. We measure precision, recall and the
F1 score, which are defined as follows
 $$\mbox{precision}=\frac{tp}{tp+fp}$$
 $$\mbox{recall}=\frac{tp}{tp+fn}$$
$$\mbox{F1 score}=2*\frac{\mbox{precision*recall}}{\mbox{precision+recall}}$$

\begin{figure}[!t]
\centering
\includegraphics[width=\linewidth]{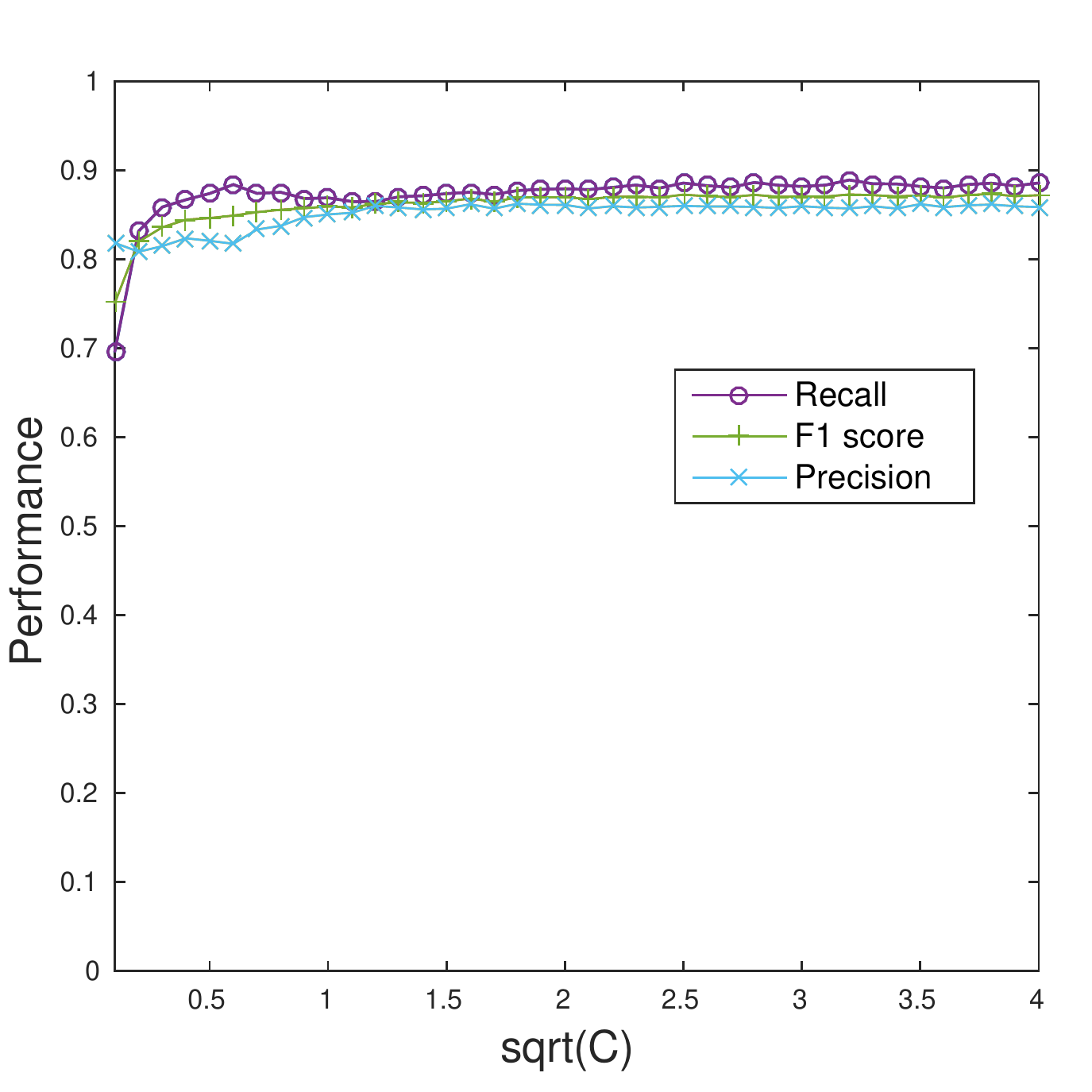}
\caption{Perfomance of the SVM classifier}
\label{fig:svm}
\end{figure}

\begin{figure}[!t]
\centering
\includegraphics[width=\linewidth]{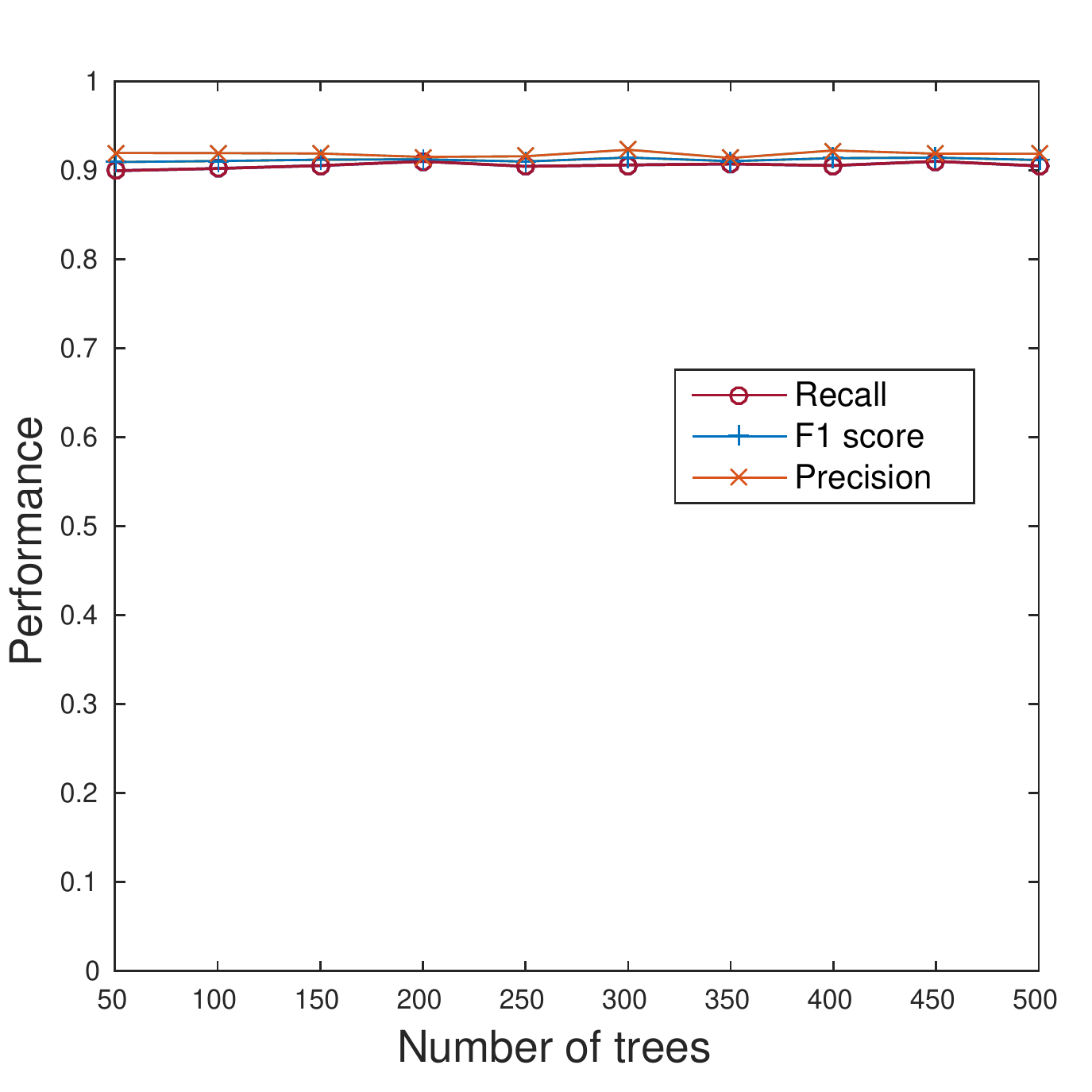}
\caption{Perfomance of the bagged decision trees}
\label{fig:rf}
\end{figure}

The models were trained with varying
values of certain parameters. In the case of support vector machines, we
employed varying values of the parameter $C$ in equation
\ref{eq:svm_primal}. Figure \ref{fig:svm} shows how the model behaves
as we increase this value. As $C$ approaches infinity, the model
becomes closer to the hard-margin variant and the training process
becomes more expensive. It can be seen that no significant gains are
obtained beyond a value of $C=2^2$ (note the root scale for
$C$). At a certain point, the model is expected to overfit the
training data
and thus the errors on test data will increase. In the case of random
forests, we train models using different 
numbers of decision trees. Figure \ref{fig:rf} shows how the
performance varies as more trees are added to the ensemble. It can be
seen that no significant improvement is gained from adding more than
50 trees.

Table \ref{tab:results} shows the values of
the three performance metrics for the best models in terms of F1 score
for both SVMs and random forests. The best SVM model was trained with
$C=3.8^2$. However, the edge over more lenient and thus more
efficiently trainedd models is very moderate. The best random forest
was trained with 300 trees.

\begin{table}[!t]
\centering
\begin{tabular}{|c|c|c|}
\hline
\multicolumn{3}{|c|}{Classification performance} \\
\hline
& SVM (C=$3.8^2$) & Random forests (300 trees) \\
\hline
Precision & 0.8616 & 0.9232 
\\ \hline
Recall & 0.8856 & 0.9061
\\ \hline
F1 score & 0.8734 & 0.9144
\\ \hline

\end{tabular}
\vspace{1em}
\caption{Performance of the best models.}
\label{tab:results}
\end{table}

\section{Conclusions \& future work}
\label{sec:conclusion}
In this paper we have studied how machine learning can help in managing
NFV infrastructure, which is expected to become a prevalent approach
in 5G networks. Specifically, we have shown that standard models for
classification can detect the noisy neighbor phenomenon with an
accuracy of more than 90\% in a simple scenario. In the future it
would be interesting to evaluate how these models perform in more
complex scenarios. It would also be interesting to extract more
features from the infrastructure as well as to employ more
sophisticated models for improved classification rates.


\section*{Acknowledgment}
The research leading to these results has received funding from the
    European Union under the H2020 grant agreement n. 671625 (project
    CogNet).

The authors would like to thank Ignacio Rubio L\'{o}pez for his valuable help
running experiments.



%

\bibliography{IEEEabrv,noisy}{}
\bibliographystyle{plain}

\end{document}